\documentclass[graybox]{svmult}

\usepackage{mathptmx}       
\usepackage{helvet}         
\usepackage{courier}        
\usepackage{type1cm}        
\usepackage{makeidx}         
\usepackage{graphicx}        
\usepackage{multicol}        
\usepackage[bottom]{footmisc}

\usepackage{bm}   
\usepackage{amsmath}
\usepackage{amssymb}

\makeindex             

\begin{document}

\title*{Mechanics of Apparent Horizon in Two Dimensional Dilaton Gravity}
\author{Rong-Gen Cai and Li-Ming Cao}
\institute{Rong-Gen Cai \at Institute of Theoretical Physics, Chinese
Academy of Sciences, P.O. Box 2735, Beijing  100190, China, \email{cairg@itp.ac.cn}
\and Li-Ming Cao \at Interdisciplinary Center for Theoretical Study,
 University of Science and Technology of China, Hefei, Anhui 230026, China, \email{caolm@ustc.edu.cn}}
%
%
\maketitle

\abstract*{Each chapter should be preceded by an abstract (10--15 lines long) that summarizes the content. The abstract will appear \textit{online} at \url{www.SpringerLink.com} and be available with unrestricted access. This allows unregistered users to read the abstract as a teaser for the complete chapter. As a general rule the abstracts will not appear in the printed version of your book unless it is the style of your particular book or that of the series to which your book belongs.
Please use the 'starred' version of the new Springer \texttt{abstract} command for typesetting the text of the online abstracts (cf. source file of this chapter template \texttt{abstract}) and include them with the source files of your manuscript. Use the plain \texttt{abstract} command if the abstract is also to appear in the printed version of the book.}

\abstract{In this article, we give a definition of  apparent horizon in a two dimensional general
dilaton gravity theory. With this definition, we construct the mechanics of
the apparent horizon by introducing a quasi-local energy of the theory. Our discussion
generalizes the apparent horizons  mechanics in general
spherically symmetric spactimes in four or higher dimensions to the two dimensional dilaton gravity case.}

\section{Introduction}
\label{sec:1}
Quantum theory together with general relativity predicts that black
hole behaves like a black body, emitting thermal radiation, with a
temperature proportional to the surface gravity of the  black hole
 and with an entropy proportional to the area of the cross section
of the event horizon
\cite{Haw,Bek}. The Hawking temperature and Bekenstein-Hawking entropy together
with the black hole mass obey the first law of
thermodynamics~\cite{firstlaw}. 
The first law of thermodynamics of
black hole has two different versions---phase space version or
passive version and  physical process or active version~\cite{Wald}.
In these two versions of discussion, the stationary of the black hole 
is essential, and the discussion is focused on the event horizon of these stationary spacetimes.
However, this kind of
horizon strongly depends on the global structure of the spacetime and there exist  some practical issues which can not be easily
solved~\cite{Ashtekar}.   It is very interesting to note that gravitational field equation on the black hole horizon can be expressed 
into a first law form of thermodynamics~\cite{Pad,Pad1,Cai-Ohta}.
The usual approach to  black hole
thermodynamics is  to start  with  the dynamics of gravity and ends with  the thermodynamic of
black hole spacetimes. Can we  turn the logic around and get the dynamics of gravity from some thermodynamic considerations?
This can not be fulfilled in this traditional approach because of the dependence of the global spacetime information
of the event horizon.
 Apparent horizon defined by Hawking does not rely on the causal structure of the spacetime. However,  it still depend on some
global information of the spacetime---one has to select a slicing of the spacetime in advance. Furthermore, it is not clear how to
establish thermodynamics on general apparent horizons.  To reveal the relation between the spacetime dynamics and thermodynamics, 
probably local or quasilocal defined horizons are necessary.

A local or quasi-local definition of
horizon is based on the local geometry of the spacetime. So it has a potential possibility to provide us  more hints
to study the relation between some fundamental thermodynamics  and  the gravitational equations. Along
this way,  fruitful results have been obtained. In fact, based on
local Rindler horizon, Jacobson {\it et al} \cite{Jac, Jac1} was
able to derive gravity field equation from the fundamental Clausius
relation. With the assumption of FRW spacetime, Cai  and Kim have obtained the 
Friedmann equations from the fundamental thermodynamical relation $\mathrm{d}E=T\mathrm{d}S$ on the apparent horizon
of the spacetimes~\cite{Cai:2005ra}. A simple summary on the relation between the spacetime dynamics and thermodynamical first law can be 
found in Ref.~\cite{Cai,Cai1}, while the further understandings of gravitational dynamics from thermodynamical aspects can be seen in Ref.~\cite{Pad2}.

On the other hand, for general dynamical black holes,
Hayward has proposed a new horizon,  trapping horizon, to study
associated thermodynamics in $4$-dimensional Einstein
theory without the stationary assumption~\cite{ Hayward2}. In this theory, for general spherically symmetric spacetimes,
Einstein equations can be rewritten in a form called ``unified first
law". Projecting this unified first law along trapping horizon, one
gets the first law of thermodynamics for dynamical black holes. This trapping
horizon can be null,  spacelike, and timelike, and has no direct relation to the causal
structure of the spacetime. Inspired by this
quasilocal definition of  horizon, Ashtekar {\it et al}  have proposed two
types of  horizons, i.e.,  isolated horizon and dynamical horizon.
The former is null, while the later is spacelike~\cite{Ashtekar}. The mechanics of
these horizons also has  been constructed.
In some sense, the trapping horizon is a generalization of the Hawking's apparent horizon.
However, the slicing of the spacetimes is not necessary to define this horizon.  Nevertheless, in
this paper, we still use the terminology  of apparent horizon.  Of course, it has the same meanings as
the trapping horizon, and can be understood as a generalized apparent horizon.

Two dimensional dilation gravity theory has been widely studied over the past twenty years.
One can get this kind of gravity from the spherically symmetric reduction of Einstein gravity theory in higher dimensions.
 To eliminate  Weyl anomaly on string
world sheet, one  also has such a kind of gravity theory, for example, the
famous CGHS model and others, for a nice review see~\cite{Grumiller:2002nm}. In these theories, a lot of black hole
solutions and cosmological solutions have been found. When some matter fields are included, in general the situations become  complicated, 
and usually we have to study general dynamical
solutions. In these two dimensional dilaton gravity theories,
the apparent horizon has been used  for a long time.  However, what is the meaning
of the apparent horizon in a two dimensional  theory? Obviously,  apparent horizon
can not be defined in the usual way because we can not define any expansion scalar of  a null congruence in two dimensional spacetimes.
In other words, the codimension-2 surface shrinks to a point in a two dimensional spacetime, and
intuitively, the size of the point does not change along a light-like geodesic,  so the expansion does not make any sense. 
In this paper, we will propose a definition of apparent horizon in these two dimensional general dilaton gravity theories.
With this definition in hand, we can construct the mechanics of the apparent horizon by introducing a quasilocal energy.
This energy is similar to the Misner-Sharp energy in four or higher dimensional Einstein gravity theory. Actually, it can be found that
this energy reduces to the usual Misner-Sharp energy for a special kind of  dilaton gravity theory  coming from the spherical reduction
of higher dimensional Einstein gravity theory.

\section{General dilaton gravity theory in two dimensions}
\label{sec:2}
For a general two dimensional dilaton gravity theory, its action can be written into a  following form~\cite{Grumiller:2002nm}
\begin{equation}
I=\int d^2x\sqrt{-h}\left[\Phi R+U(\Phi)D^a\Phi D_a\Phi+V(\Phi)+
\mathcal{L}_m\right]\, ,
\end{equation}
where $\Phi$ is the so-called dilaton field, and $R$ is the two dimensional Ricci
scalar. The matter Lagrangian is represented by  $\mathcal{L}_m$  which may contain tachyon (and others). The matter is
 denoted by $\psi$ for simplicity. So, in general, the matter Lagrangian can be expressed as
$$\mathcal{L}_m= \mathcal{L}_m(\psi, D_a\psi\, ,\cdots \, ,\Phi\, , D_a\Phi\, ,\cdots)\, .$$ The equation of motion for the dilaton field $\Phi$ can be
written as
\begin{equation}
R-U'(\Phi)D^a\Phi
D_a\Phi+V'(\Phi)-2U(\Phi)\Box \Phi+\mathcal{T}_m=0\, ,
\end{equation}
where $\Box = D_aD^a$ and the prime  stands for the derivative with respect to $\Phi$:  $\mathrm{d}/\mathrm{d}\Phi$, while the scalar $\mathcal{T}_m$
is defined as
 $$ \mathcal{T}_m=\frac{\partial \mathcal{L}_m}{\partial\Phi}-D_a\frac{\partial \mathcal{L}_m}{\partial D_a\Phi}+\cdots \, .$$
Of course, if the dilation field does not  couple to the matter field,
this term vanishes. The Euler-Lagrangian equation for the matter field $\psi$ can be obtained in a similar way.
The equations of motion for the metric $h_{ab}$
can be put into a form
\begin{eqnarray}
\label{eommetric} U(\Phi)D_a\Phi D_b\Phi-\frac{1}{2}U(\Phi)D^c\Phi
D_c\Phi h_{ab} -D_{a}D_{b}\Phi \nonumber\\
+\Box \Phi
h_{ab}-\frac{1}{2}V(\Phi)h_{ab}= T_{ab}\, ,
\end{eqnarray}
where  $T_{ab}$ is the energy-momentum tensor of the matter field. Straightforward calculation shows the covariant divergence of this energymomentum tensor is given by
\begin{eqnarray}
D^aT_{ab}=-\frac{1}{2}\big[R-U'(\Phi)D^c\Phi
D_c\Phi+V'(\Phi)
-2U(\Phi)\Box\Phi\big]D_b\Phi\, ,
\end{eqnarray}
and this suggests that
\begin{equation}
\label{DaTab}
D^aT_{ab}=\frac{1}{2}\mathcal{T}_m D_b\Phi\, .
\end{equation}
Thus we see that the dilation provides an external force to the matter field when the coupling between the matter field and dilaton is present.

\section{Some solutions of the theory}
\label{sec:3}
When the matter field  is absent,  we have $T_{ab}=0$ and $\mathcal{T}_m=0$.  In
Eddington-Finkelstein gauge, the general solution of Eqs.(\ref{eommetric})
 has a simple form~\cite{Grumiller:2002nm}
\begin{equation}
\label{Eddington-Finkelstein}
h=e^{Q}\left[2\mathrm{d}v\mathrm{d}\Phi-(w-2\mathfrak{m})\mathrm{d}v^2\right]\, .
\end{equation}
Here, two functions $Q(\Phi)$ and $w(\Phi)$ have been introduced, and
they are defined by
\begin{equation}
\label{UVdefinition} U=-Q'\, ,\quad V=e^{-Q}w'
\end{equation}
up to some constants. Note that $e^Q\mathrm{d}\Phi$ is a closed form,  from Poincare lemma,  there exists a  function $r$ satisfying $\mathrm{d}r=e^Q\mathrm{d}\Phi$.
This means the general solution of the metric can be transformed into a familar form
\begin{equation}
\label{staticmetric}
h=\left[2\mathrm{d}v\mathrm{d}r-e^Q(w-2\mathfrak{m})\mathrm{d}v^2\right]\, .
\end{equation}
One can
replace the constant $\mathfrak{m}$  by a function of $v$, i.e.,
$\mathfrak{m}(v)$, and then construct a typical dynamical spacetime,
i.e., Vaidya-like spacetime.  In that case,  the energy-momentum tensor of matter field  will no longer vanish, and has a nontrivial component $T_{vv}$  satisfying
\begin{equation}
\label{Bondiequation}
\frac{\mathrm{d}\mathfrak{m}(v)}{\mathrm{d}v}=T_{vv}\, .
\end{equation}
Naively one can read
off the apparent horizon of this spacetime in this
Eddington-Finkelstein gauge---it is given by equation
\begin{equation}
\label{withoutmatter}
w-2 \mathfrak{m}(v)=0=e^{Q}D_a\Phi D^a\Phi\, .
\end{equation}
However, we may ask a question here --- What is the apparent horizon in this two dimensional spacetime? In the above discussion, we
have read off naively the location of apparent horizon from the metric in Eddington-Finkelstein gauge.  But in what sense it is an apparent horizon? Usually, the definition
of apparent horizon depends on the extrinsic geometry of codimenstion-2 spacelike surface, i.e., the expansion scalars of the surface. Now we are considering two dimensional spacetime,
the codimension-2 surface shrinks to a point,  and the expansion scalars can not be defined.  In the next section, we will focus on this question, and explain why the location of $D_a\Phi D^a\Phi=0$ can be viewed as
the apparent horizon in the two dimensional case.

\section{Apparent horizon}
\label{sec:4}

In this section, we define the  apparent horizons in the two dimensional  spacetime of the dilaton gravity theory.
Assume  $\{\ell^a, n^a\}$ is a null frame in the spacetime, and the metric can be expressed as
\begin{equation}
h_{ab}=-\ell_an_b - n_a\ell_b\, ,
\end{equation}
where $\ell^a$ and $n^a$ are two null vector fields which are globally defined on the spacetime and satisfy $\ell_an^a=-1$.
We assume $\ell^a$ and $n^a$ are both future pointing, and furthermore, $\ell^a$ and $n^a$ are outer pointing and inner pointing respectively.
On the spacetime, there is a natural vector field $\phi^a=D^a\Phi$. Obviously, the causality of the vector field is determined by the signature of $\phi_a\phi^a$.
According to the causality of this vector, the spacetime can be divided into several parts, and in each part the vector field $\phi^a$ either spacelike or timelike.
$\phi^a$ is null on the boundary of any part, and this boundary can be defined as a kind of horizon. It is easy to find  that on this horizon we have
$$\phi_a\phi^a=D_a\Phi D^a\Phi
= -2\mathcal{L}_{\ell}\Phi \mathcal{L}_{n}\Phi=0\, .$$
So on these horizons we have $\mathcal{L}_{\ell}\Phi=0$ or $\mathcal{L}_n\Phi=0$. We can further classify the horizons as follows.
The horizon is called future if $\mathcal{L}_{\ell}\Phi =0$ and $\mathcal{L}_n\Phi < 0$. In this case, if
$\mathcal{L}_n\mathcal{L}_{\ell}\Phi < 0$, we call the horizon is outer. The future horizon with $\mathcal{L}_n\mathcal{L}_{\ell}\Phi > 0$ is called inner.
The past horizon is defined by $\mathcal{L}_n\Phi = 0$, and $\mathcal{L}_{\ell}\Phi > 0$. Similarly, the past
horizon with $\mathcal{L}_{\ell}\mathcal{L}_n \Phi > 0$ is called outer, and the case with $\mathcal{L}_{\ell}\mathcal{L}_n\Phi < 0$
is called inner. Mimicking the cases in higher dimensions, the region with $\mathcal{L}_{\ell}\Phi <0$ and $\mathcal{L}_{n}\Phi <0$ (or $\phi_a\phi^a<0$) can be called trapped region of the spacetime~\cite{Hayward2}.

At the first sight, these definitions have nothing to do with the geometry of the spacetime. How do these definitions realize the description
of the spactime region where light can not escape? To answer this question, we have to investigate the detailed structure of the definition
and the equations of motion of the dilaton theory. Some calculation shows
\begin{eqnarray}
\label{LLPhi}
\mathcal{L}_{n}\mathcal{L}_{\ell}\Phi &=& -\kappa_{(n)}(\mathcal{L}_{\ell}\Phi) -(1/2) \Box \Phi\, ,\nonumber\\
\mathcal{L}_{\ell}\mathcal{L}_{n}\Phi &=& -\kappa_{(\ell)}(\mathcal{L}_{n}\Phi )-(1/2) \Box \Phi\, ,\nonumber\\
\mathcal{L}_{\ell}\mathcal{L}_{\ell}\Phi&=&\kappa_{(\ell)} (\mathcal{L}_{\ell}\Phi) + U(\Phi)(\mathcal{L}_{\ell}\Phi)^2 - T_{ab}\ell^a\ell^b\, ,\nonumber\\
\mathcal{L}_{n}\mathcal{L}_{n}\Phi&=&\kappa_{(n)} (\mathcal{L}_{n}\Phi) + U(\Phi)(\mathcal{L}_{n}\Phi)^2 - T_{ab}n^an^b\, ,
\end{eqnarray}
where we have used the equation (\ref{eommetric})  and introduced two scalars
$\kappa_{(\ell)}=-n_a\ell^bD_b\ell^a$, $\kappa_{(n)}=-\ell_an^bD_bn^a$.
From these equations, it is easy to find that on the future outer horizon, we have
\begin{equation}
\label{BoxPhi}
\Box \Phi > 0\, .
\end{equation}
 Similarly, on the past outer horizon, we have
$\Box \Phi<0$. In the following discussion, we will focus on the future outer horizon.

Here, we give some explanation why we can define the horizon in this way. From Eqs.(\ref{LLPhi}), we have
\begin{eqnarray}
\label{LkPP}
\mathcal{L}_{k}(\phi_a\phi^a)=  \alpha \Big[ \Box \Phi (\mathcal{L}_{\ell}\Phi)+ 2 ( T_{ab}\ell^a\ell^b ) (\mathcal{L}_{n}\Phi)
- 2U(\Phi)(\mathcal{L}_{n}\Phi) (\mathcal{L}_{\ell}\Phi)^2\Big]\, ,
\end{eqnarray}
where $k^a=\alpha \ell^a$ is some null vector field and $\alpha$ is a positive function such that the parameter of $k^a$ is affine. We consider the region very near the  future outer horizon where $\mathcal{L}_{\ell}\Phi=0$ and $\mathcal{L}_n\Phi <0$. In this small neighbourhood, from continuity, $\Box \Phi$ should be positive and $\mathcal{L}_{\ell}\Phi$ is very small. Now let us consider the part of the neighbourhood inside the trapped region of the spacetime (where $\mathcal{L}_{\ell}\Phi$ is a small negative  quantity and $\mathcal{L}_n\Phi$ is a finite negative quantity). In this case, we have
\begin{equation}
\mathcal{L}_{k}\|\phi\|<0 \, ,\qquad \|\phi\|=\sqrt{|\phi_a\phi^a|}\, ,
\end{equation}
only when the null energy condition is broken, i.e., $T_{ab}\ell^a\ell^b<0$. This mathematical relation suggests the light with wave vector $k^a$ can approach the
line with $\|\phi\|=0$ (inside the trapped region) only when the null energy condition of the matter field is broken. This can not happen for usual classical matter field.
So the outward propagating light do not exist near the future outer horizon. Similarly, we have
\begin{eqnarray}
\label{LkPP1}
\mathcal{L}_{k}(\phi_a\phi^a)= \alpha \Big[ \Box \Phi (\mathcal{L}_{n}\Phi)+ 2 ( T_{ab}n^an^b ) (\mathcal{L}_{\ell}\Phi)
- 2U(\Phi)(\mathcal{L}_{n}\Phi)^2 (\mathcal{L}_{\ell}\Phi)\Big]\, ,
\end{eqnarray}
where $\alpha$ is still a positive function. Obviously, inside the trapped region and near the future outer horizon, we have $\mathcal{L}_k\|\phi\|>0$. This means
that inward propagating light is always allowed whatever the energy condition is satisfied or not.

Now, let us consider the neighborhood of the horizon inside the region where $\phi_a\phi^a>0$.  In the region,  $\mathcal{L}_{\ell}\Phi $ is a small positive quantity.
From Eq.(\ref{LkPP}), it is easy to find that it is possible to get $\mathcal{L}_k\|\phi\|>0$ in this case especially when matter field is absent. This means the light has possibility to
escape from this region to the region with large value of $\|\phi\|$. For the  inner pointing light, Eq.(\ref{LkPP1}) suggests it can cross the horizon and can reach the trapped region.

In a word, light cannot escape from the trapped region we have defined. So the horizon we have defined has the same properties as the apparent horizon in higher dimensions which, roughly speaking,
can be viewed as the boundary of trapped region.
So in this paper we still use the terminology of apparent horizon to describe this kind of horizon.
In the next section, by introducing a quasilocal energy  in this dilaton gravity theory, the mechanics of the apparent horizon will be established.

\section{The mechanics of the apparent horizon}
\label{sec:5}

To study the mechanics of the apparent horizon, we have to define the
quasilocal energy inside the horizon. Generally, this is not an easy task.
However, in the dilaton gravity we are considering, there is a well defined
quasilocal energy. This can be found as follows.
From the energy-mometum tensor of the matter field, we can define two useful quantities , i.e., a scalar called generalized pressure
\begin{equation}
P=-\frac{1}{2}T^a{}_a\, ,
\end{equation}
and a vector called energy-supply,
\begin{equation}
\Psi_a=T_a{}^be^{Q}D_b\Phi+Pe^{Q}D_a\Phi\, .
\end{equation}
It is easy to find
\begin{eqnarray}
\Psi_a=e^{Q}\bigg{[}\frac{1}{2}U(\Phi)D^c\Phi D_c\Phi
D_a\Phi-D_aD^b\Phi D_b\Phi
+\frac{1}{2}D^cD_c\Phi D_a\Phi\bigg{]}\, .
\end{eqnarray}
Thus we have
\begin{eqnarray}
\label{APsidW} \Psi_a+Pe^QD_a\Phi=\frac{1}{2}e^{Q}\bigg{[}U(\Phi)(D^c \Phi
D_c \Phi) D_a\Phi\nonumber\\
-D_a(D^c\Phi D_c \Phi)+V( \Phi )D_a \Phi \bigg{]}\, .
\end{eqnarray}
It is not hard to find that the right hand side of the equation
(\ref{APsidW}) can be written as
\begin{equation}
 \frac{1}{2}D_a\left[w\left(1-\frac{e^Q}{w}D^c\Phi
D_c\Phi\right)\right]\, .
\end{equation}
In the above equation, $Q$ and $w$ are the same as those given in Eq.(\ref{UVdefinition}).
Comparing with the unified first law in higher dimensional
spherical symmetric spacetime~\cite{Hayward2},
we can define a
similar quasi-local energy
\begin{equation}
\label{energy} E=\frac{1}{2}\left[w\left(1-\frac{e^Q}{w}D^c\Phi
D_c\Phi\right)\right]\, .
\end{equation}
Then the equations of motion for the metric, i.e., equation
(\ref{eommetric}) can be put into the form
\begin{equation}
D_aE=\Psi_a+P e^QD_a\Phi\, .
\end{equation}
Since $e^{Q}D_a\Phi$ is a closed one form in the spacetime, at least locally,  it can be expressed as $e^QD_a\Phi = D_a\mathcal{V} $ for some function $\mathcal{V}$. This suggests the above relation can be transformed into a simple form, i.e.,
\begin{equation}
\label{unifieldfirstlaw}
\mathrm{d}E=\Psi+P \mathrm{d}\mathcal{V}\, .
\end{equation}
This energy $E$ generalizes the Misner-Sharp energy in higher
dimensional theory to the two dimensional dilaton gravity theory, and at the same time, the above equation establishes  the unified first
law in this two dimensional gravity theory.

Now let us consider the special case where the matter  field is absent, i.e.,
$T_{ab}$ is vanishing, so do $P$ and $\Psi$.  From the above unified
first law (\ref{unifieldfirstlaw}), we have $\mathrm{d}E=0$. This means that
$E$ is a constant which can be denoted by  $\mathfrak{m}$. By this consideration,  from
the energy form (\ref{energy}), we have Eq.(\ref{withoutmatter}) with  $\mathfrak{m}(v)$.
When the energy-momentum tensor  is given by some radiation matter, the general solution is just the
Vaidya-like spacetime mentioned in the  previous section. In this case, it is easy to find
that $E$ is nothing but $\mathfrak{m}(v)$ and the first law (\ref{unifieldfirstlaw}) reduces to
the Bondi's  energy balance equation (\ref{Bondiequation}).

Assume on the apparent horizon, i.e., on the spacetime points which
satisfy  $D^c\Phi D_c\Phi=0$, that the dilaton field $\Phi$
takes value $\Phi_A$, then, on the apparent horizon, the quasi-local energy becomes
\begin{equation}
E=\frac{1}{2}w(\Phi_A)\, .
\end{equation}
In general, this energy is not constant because  $\Phi_A$ may
depend on the coordinates. For example, for the Vaidya-like
spacetime, the total energy inside the apparent horizon is
$\frac{1}{2}w(\Phi_A)=\mathfrak{m}(v)$. For the static case without
matter, the apparent horizon coincides with the event horizon (if it can be defined), this
energy becomes
$\frac{1}{2}w(\Phi_A)=\frac{1}{2}w(\Phi_+)=\mathfrak{m}$, where
$\Phi_+$ is the value of dilation on the event horizon.

On the apparent horizon, the energy-supply becomes
\begin{equation}
\Psi_a=\frac{1}{2}e^{Q}\bigg{[}-D_a(D^c\Phi D_c\Phi)+\Box\Phi
D_a\Phi\bigg{]}\, .
\end{equation}
Let $\xi$ be the vector tangent to the apparent horizon. Since
$D^c\Phi D_c\Phi$ is a constant on the apparent horizon,
$\xi^aD_a(D^c\Phi D_c\Phi)=0$,  then we find
\begin{equation}
\xi^a\Psi_a=\frac{1}{2}e^{Q}(\Box\Phi) \mathcal{L}_{\xi}\Phi\, ,
\end{equation}
where $\mathcal{L}_{\xi}$ is Lie derivative along the vector $\xi$. In higher dimensions, the surface gravity of an apparent horizon
 is defined by the Kodama vector field $K^bD_{[b}K_{a]}=\kappa K_a$~\cite{Maeda:2007uu}. Here we can also
introduce a Kodama-like vector field as $K^a=-e^{Q}\epsilon^{ab}D_b\Phi$. Some calculation shows
\begin{equation}
K^bD_{[b}K_{a]}= \frac{1}{2}\Big[e^Q \Box \Phi K_a - U(e^Q D_b\Phi D^b\Phi) K_a\Big]\, .
\end{equation}
From the definition of the surface gravity,  it is easy to find  the surface gravity of the apparent horizon
can be expressed as $$\kappa=\frac{1}{2}e^{Q}(\Box\Phi)\, .$$  On the future outer apparent horizon, from Eq.(\ref{BoxPhi}), we see this surface gravity is always positive.
Therefore, we find that
the energy-supply projecting onto the apparent horizon gives
\begin{equation}
\xi^a\Psi_a=\kappa~ \mathcal{L}_{\xi}\Phi\, .
\end{equation}
As a result, on the apparent horizon, we have a relation
\begin{equation}
\mathcal{L}_{\xi}E=\frac{\kappa}{2\pi}
\mathcal{L}_{\xi}S+P\mathcal{L}_{\xi}\mathcal{V}\, ,
\end{equation}
where $S=2\pi \Phi$. This relation is the same as the first law of thermodynamics if we identify $T=\kappa/2\pi$
and regard  $S$ as entropy.  Actually, the ``entropy" of the future outer apparent horizon can also be written as
\begin{equation}
\label{entropy} S=2\pi\Phi_A\, .
\end{equation}
In general, this entropy is not a constant, and it might change with some coordinate.
In the static case, the future outer apparent horizon coincides with the event
horizon, this entropy becomes $S=2\pi\Phi_+$, this result has been
found in many  static black holes in the two dimensional dilaton gravity theories~\cite{Grumiller:2002nm}.

The function $\mathcal{V}$ can be viewed  as a kind of ``volume" of the system. It comes from the divergence free of the Kodama
vector $K^a$, i.e., $D_aK^a=0$, and can be viewed as a conserved quantity of the theory. Besides the function $\mathcal{V}$, the energy $E$ can also be viewed as a
conserved quantity.  Actually, from Eqs. (\ref{eommetric}) and (\ref{DaTab}), one can prove that $J^a=T^{a}{}_{b}K^b$ is conserved, i.e., we have $D_aJ^a=0$. This suggests that the Hodge
dual of $J_a$ is a closed one form, and locally it is the exterior derivative of a function. This function is nothing but the energy $E$.  Such kind of discussion can also be found in \cite{Mann:1992yv} where
some special matter Lagrangian has been considered.

For the  case of the energy-momentum tensor given by conformal
matter, for example, tachyon,  the trace of the energy-momentum tensor
vanishes. In this case, from the trace part of Eq. (\ref{eommetric}), the
surface gravity becomes
$
\kappa=-\frac{1}{2}w'(\Phi_A)\, .
$
This is very similar to the static case where the surface gravity is just
given by $-\frac{1}{2} w'(\Phi_+)$. The work term vanishes due to
the traceless of the energy-momentum tensor, so the first law on the
apparent horizon  becomes
\begin{equation}
\mathcal{L}_{\xi}E=\frac{\kappa}{2\pi} \mathcal{L}_{\xi}S\, .
\end{equation}
This means for the system with conformal matter, there is no
external work term. The above result shows that,  in the case with conformal matter, all the thermodynamic
quantities of the apparent horizon can be obtained from the static  case through the
replacement of  $\Phi_+$ by $\Phi_A$.

For the two dimensional dilation gravity coming from the 
spherically symmetric reduction of $n$-dimensional Einstein gravity, the potential $U$ and $V$ are given
respectively by
\begin{equation}
\label{UVfuction} U(\Phi)=\frac{n-3}{n-2}\Phi^{-1}\, ,\quad
V(\Phi)=(n-2)(n-3)\lambda^2\Phi^{\frac{n-4}{n-2}},
\end{equation}
where $\lambda$ is a constant with dimension of mass square, and
$\Phi=(\lambda r)^{n-2}$.  If the function $U$ and $V$ have the forms
(\ref{UVfuction}), the quasi-local energy (\ref{energy}) is just the
$n$-dimensional Misner-Sharp energy
\begin{equation}
E_{MS}=\frac{1}{2}(n-2)r^{n-3}\left(1-D^ar D_ar\right)\, .
\end{equation}
It is straightforward to see that the entropy of apparent horizon in
the two dimensional dilaton gravity is given by the area of the horizon sphere in $n$-dimensions. This
can be obtained  by replacing  $\Phi_A$ in equation (\ref{entropy}) by
$(\lambda r_A)^{n-2}$. It should be noted here that our discussion is not
restricted to the future outer apparent horizon. Actually, the same discussions
can be applied to other types of  apparent horizons. For example, in an FRW
universe, the entropy of the apparent horizon is given by the one quarter of the  area of
the apparent horizon, and the radius of the apparent horizon can be
expressed by Hubble parameter as
$$
\frac{1}{\tilde{r}_A^2}= H^2+\frac{k}{a^2}\, ,
$$
where $\tilde{r}_A=ar_A$, and $a$ is the scale factor in the FRW
universe, see for example~\cite{Cai:2006rs}. Thus the apparent horizon associated with the 
FRW universe also applies here. 

\begin{acknowledgement}
This work  is dedicated to celebrate the 60th birthday of Prof. T. Padmanabhan.
The work was supported in part by the National Natural Science Foundation of China with grants
No.11205148, No.11235010, No.11375247 and No.11435006.
\end{acknowledgement}
%
%

\end{document}